\documentclass{aa}  

\usepackage{graphicx}
\usepackage{txfonts}

\usepackage{color}

\begin{document}

   \title{Spatial distribution of jets in solar active regions}


     \author{J. Odermatt  \inst{1}
          \and
          K. Barczynski \inst{1,2}
          \and
          L.K. Harra \inst{2,1}
          \and
          C. Schwanitz \inst{1,2}
          \and
          S\"am Krucker \inst{3,4}
          }

   \institute{ ETH-Zurich, H\"onggerberg campus, HIT building, Z\"urich, Switzerland \\
              \email{krzysztof.barczynski@pmodwrc.ch}
         \and
            PMOD/WRC, Dorfstrasse 33, CH-7260 Davos Dorf, Switzerland
      \and
        University of Applied Sciences and Arts Northwestern Switzerland, Bahnhofstrasse 6, 5210 Windisch, Switzerland
        \and
        Space Science Laboratory, University of California, \\
       Berkeley, CA 94720-7450, USA
              }

   \date{Received Month Day, Year; accepted Month Day, Year}

 
  \abstract
   {Solar active regions are known to have jets. These jets are associated with heating and the release of particles into the solar wind. }
   {Our aim is to understand the spatial distribution of coronal jets within active regions to understand if there is a preferential location for them to occur. }
   {We analysed five active regions using Solar Dynamics Observatory Atmospheric Imaging Assembly data over a period of 2-3.5 days when the active regions were close to disk centre. Each active region had a different age, magnetic field strength, and topology. We developed a methodology for determining the position and length of the jets. }
   {Jets are observed more frequently at the edges of the active regions and are more densely located around a strong leading sunspot. The number of coronal jets for our active regions is dependent on the age of the active region. The older active regions produce more jets than younger ones. Jets were observed dominantly at the edges of the active regions, and not as frequently in the centre. The number of jets is independent of the average unsigned magnetic field and total flux density in the whole active region. The jets are located around the edges of the strong leading sunspot.}
   {}

   \maketitle
%

\section{Introduction}

  Solar coronal jets are dynamic events that can occur in all parts of the solar atmosphere \citep[e.g.][]{Sterling_2017}. They appear as collimated flows of plasma that can extend far into the outer layers of the solar atmosphere. They occur at different spatial scales and timescales, and can take place in active regions (ARs), quiet Sun regions, and coronal hole regions irrespective of the solar cycle phase. Jets share many similarities with larger events such as solar flares and coronal mass ejections and may contribute to the heating of the solar corona and the acceleration of the solar wind \citep{Raouafi_2016}. Despite their relatively common occurrence, the underlying mechanism responsible for the production and characteristics of jets is not completely understood \citep{Raouafi_2016}. Most theories are based on the reconnection of magnetic fields. The regular appearance of X-ray jets seen by the Yohkoh mission led to the `standard jets' theory proposed by \cite{Shibata_1992}, which posits that magnetic flux emergence with subsequent reconnection with open magnetic field lines is the primary cause of coronal jets. \cite{Moore_2010} coined the new category of `blowout jets', which have a more complex topological structure and a more explosive behaviour. \cite{Sterling_2017} proposed mini-filament eruptions as a  cause for  jet occurrence. 

Much research has been carried out on jets. Since the first stereoscopic jet observation \citep{Patsourakos_2008}, temperature measurements \citep{Nistico2011}, magnetic flux cancellation analysis \citep{Panesar_2018}, and analysis of the evolution of jets \citep{Sterling_2017} and their association with particle acceleration \citep{Musset_2020} have all been undertaken.

The velocities of X-ray jets are in the range 10 to 1000 km s$^{-1}$ with a median value of 200 km s$^{-1}$ \citep{Shimojo1996}.
\citet{Joshi2017} have found velocities of recurrent jets from highly dynamic magnetic field regions in the range 47-308 km s$^{-1}$ using Solar Dynamics Observatory (SDO) Atmospheric Imaging Assembly (AIA)  171\AA~images.
However, there had been no studies regarding the spatial distribution of jets within ARs, which is what we study in this paper. We select five ARs that have different ages and magnetic field strengths to assess whether there are differences in the jet distributions.

\section{Data analysis}

To study the spatial locations of jet activity around the ARs, we analysed extreme ultraviolet (EUV) images and magnetic field data from the SDO. 

\subsection{SDO (AIA, HMI)}\label{section_SDO_AIA_HMI}
The SDO \citep{Pesnell_2012} consists of three different instruments, the AIA \citep{Lemen_2012}, the Helioseismic and Magnetic Imager \citep[HMI;][]{Schou_2012}, and the Extreme Ultraviolet Variability Experiment (EVE). For the purpose of this study, we analysed the AIA and HMI data.

The AIA instrument provides full solar disk images  (up to $0.5 R_\odot$ above the solar limb) in seven different EUV channels, two ultraviolet channels, and one visible wavelength channel. The EUV channels have a temporal and angular resolution of 12\,s and 1.2$^{\prime\prime}$, respectively. We focused on the AIA 193~\AA\ channel; it is dominated by emission from the Fe$\textsc{XII}$ emission line, which has a peak of formation temperature of log T=6.1. The AIA 193~\AA\ channel is used to determine the locations of the coronal jets, and hence we analysed only plasma at the coronal temperature. Jets are observed at all temperatures throughout the solar atmosphere.
 Multi-wavelength analysis of the jets is important and will be discussed in future studies. Because of the complexity, we decided to first focus on methodology and analyse only jets in the single channel.  

The HMI instrument provides photospheric vector magnetograms, line-of-sight (LOS) magnetograms, and Dopplergrams of the full solar disk. The LOS magnetograms have a temporal and angular resolution of $45s$ and 1.0$^{\prime\prime}$, respectively. We analysed the magnetic field strength and evolution for each AR using LOS magnetograms. The noise level for HMI LOS magnetograms is below $8G$ \citep{Yeo_2013}.

The AIA and HMI data were initially explored using JHelioviewer\footnote{\url{https://www.jhelioviewer.org}}. Then, we used pre-processed data from the Joint Science Operations Center (JSOC)\footnote{\url{http://jsoc.stanford.edu}} for more detailed analysis. The JSOC data are  corrected for plate scaling, solar rotation, and shifts (level 1.5 data). HMI data are given in units of gauss and AIA data in units of digit number (DN), which is an instrument-specific intensity-related value. We analysed five ARs, as summarised in Table~\ref{DA_table_AR_info}. We chose regions of interest located close to disk centre, [x,y]=($\pm$ 500$''$, $\pm$ 500$''$).

\subsection{Jet detection}
This section describes the detection process for the location and duration of jets in ARs.
We defined jets in a two-step process. 
The first step was a manual search on JHelioviewer using the AIA 193~\AA~filter with a 36s time interval. Any event logged for further selection fulfilled the following criteria:
(i) a collimated flow is observed; (ii) there is an increase in intensity followed by a decrease in intensity; and (iii) the duration of the intensity increase was shorter than $\sim$15min (the interval was chosen arbitrarily after the data inspection).

The lower limit of the jet observations was constrained by the time resolution of 36s in JHelioviewer. An upper limit was set by visual inspection and becomes less reliable for slow jets. 

During the processing of the data, we used the AIA~193\AA\ filter with a 12s time cadence from JSOC for the detailed data analyses. The AIA~193\AA\ images were corrected for solar rotation (using the JSOC procedure). The following additional criteria were added to this dataset: (i) the collimated flow must not be confined along a loop, to avoid structures with a magnetic field topology that can block a further expansion of the plasma; (ii) the flow must not abruptly end at a bright point; and (iii) a curved flow is allowed, if it is not confined along a closed coronal loop.

The initial visual screening resulted in a total of 530 events, of which 239 fulfilled all the required criteria to be classified as jets. We manually checked which structures fulfilled all required criteria.

\begin{table*}
\begin{tabular}{l|c|c|c|c|c|c|c|c|c}
     AR Nr. & Observation start & \begin{tabular}[c]{@{}c@{}}Obs. dur. \\ (h) \end{tabular} & \begin{tabular}[c]{@{}c@{}}Location$^{*}$ \\ (x,y) \end{tabular} & Jets&  Jets/day & \begin{tabular}[c]{@{}c@{}}AR age \\ (day) \end{tabular}  & 
     \begin{tabular}[c]{@{}c@{}}$\Phi$ \\ (Mx $\times 10^{22}$) \end{tabular}
      & \begin{tabular}[c]{@{}c@{}} <|B|$^{**}$> \\ (G) \end{tabular}&
      \begin{tabular}[c]{@{}c@{}}AR area$^{***}$ \\ (Mm$^2$) \end{tabular}\\
     \hline
     AR12737 & 00:00 UT 02 Apr 2019 & 48& (-460$^{\prime\prime}$, 270$^{\prime\prime}$) & 8 &4.0 & 3.0 & 1.1& 278 &1485 \\
     AR12738 & 00:00 UT 12 Apr 2019 & 84& (-420$^{\prime\prime}$, 170$^{\prime\prime}$) & 92 & 26.3 &23.4 & 3.7 &344&6047  \\
     AR12803   & 12:00 UT 23 Feb 2021 &48& (-241$^{\prime\prime}$, 475$^{\prime\prime}$)& 70 & 35.0& 75.8 & 1.0 &225&1327
     \\
     AR12822   & 12:00 UT 12 May  2021 & 48 & (-236$^{\prime\prime}$, 352$^{\prime\prime}$) & 50& 25.0&  18.9 & 1.5&301 & 2595 \\
     AR12823  & 12:00 UT 12 May 2021 & 48 & (-240$^{\prime\prime}$,-360$^{\prime\prime}$) & 19 & 9.5& 3.3 &0.6&256 &959 
\end{tabular}
\caption{Properties of the ARs. The AR number is given according to the National Oceanic and Atmospheric Administration (NOAA). The location is in helio-projective longitude (arcsec) and defined at the start time of the observation. The age of the AR is determined as the time from the emergence of the AR based on the AIA193\AA~images and HMI magnetographs.  The magnetic field flux ($\Phi$) is calculated only for the pixels above noise level |B|>10G.
(*) The location corresponds to the centre of the field-of-view at the start of the observation. Figures~\ref{AR_April_AIAJetflow} - \ref{AR_May_JetflowS} show the field-of-view at the middle of the observation duration. (**) The average magnetic strength is calculated only for the pixels above |B|>100G to focus only on the core of the AR. (***) The area of the AR is defined as the region where |B|>100G.}
\label{DA_table_AR_info}
\end{table*}

\subsection{Jet length}
\label{section_DA_jet_length}

We measured the length of the jets to determine if there is any variation in length with location. We measured the jet length through a ten-step procedure. This procedure aims to approximate the shape of the jet with a single curve (steps 1-6) and measure the length of the jet along this curve (steps 7-10).

Step 1: We identified a jet in the AIA~193\AA~images of the ARs based on visual inspection. We determined the approximate time where the jet is the brightest (the so-called jet time).

Step 2: We chose a sub-region of 50$^{\prime\prime}$ by 50$^{\prime\prime}$ around the jet location. If the jet exceeded this range in any of the following steps, the area was increased by 25$^{\prime\prime}$ in the respective direction. Our analysis focused on a 20-minute time series of 12-second cadence AIA 193\AA~images centred at the jet time.

Step 3: We calculated the standard deviation (STDDEV) in the time series at each spatial pixel to determine the changes in intensity with time and created a STDDEV map.

\begin{figure*}
    \includegraphics[scale=0.72]{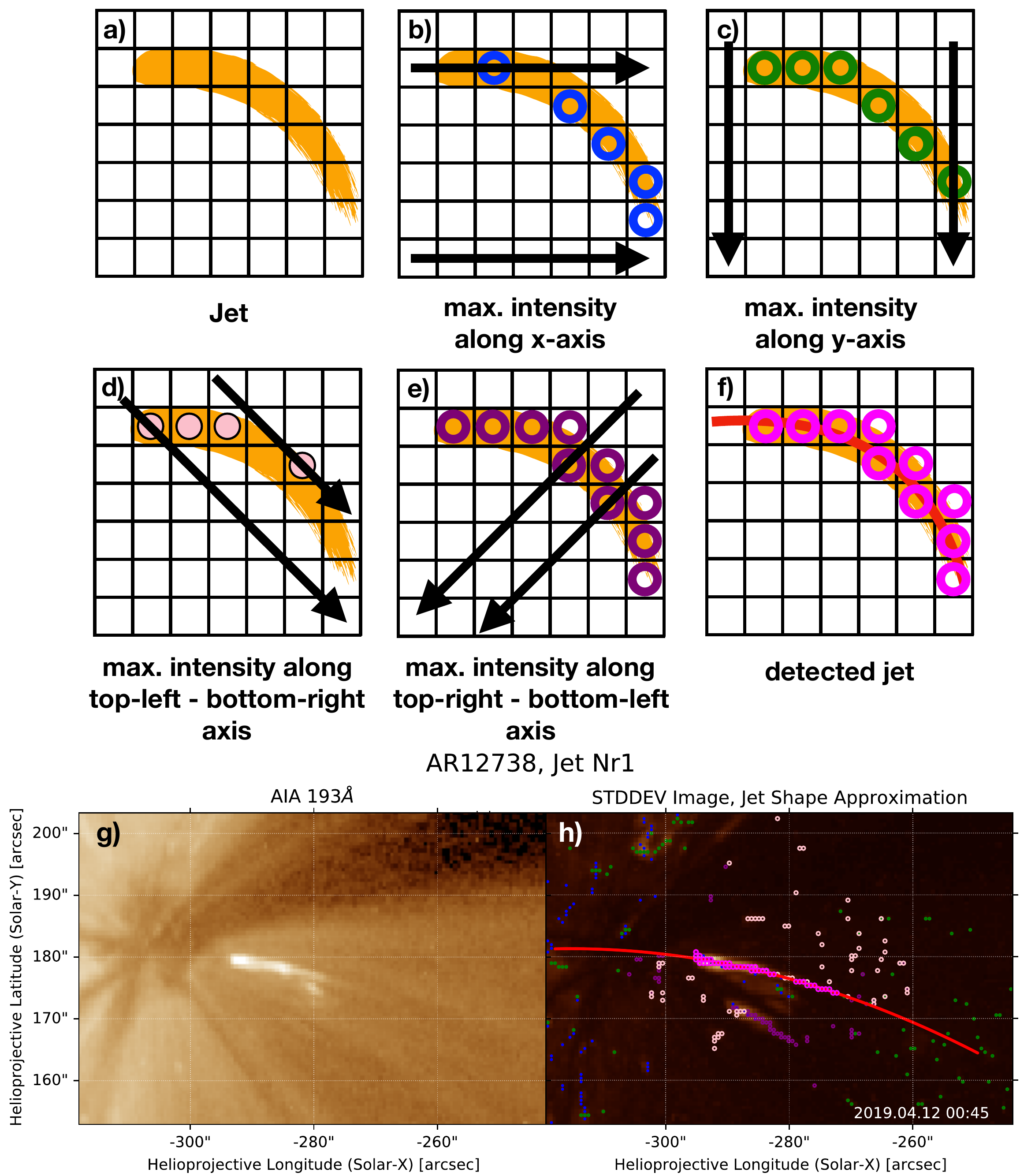}
    \caption{Jet shape determination method (panels a-f), an example jet observed in AIA 193{\AA} (panel g), and the shape determination of a jet (panel h).
    The six panels (a-f) present the schematic algorithm of the jet (yellow) shape determination.  The pixels with maximal intensities along the X and Y axis of the jet are marked in blue (panel b) and green (panel c). The pixels with maximal intensity along the diagonal axes starting from the top right and the top left are coloured in light pink (panel d) and violet (panel e). Manual selection of points is coloured in pink (panel f). The fitted quadratic polynomial is marked as a red line connecting the pink dots  (panel f).
    The jet shape (panel g) was determined by applying the method presented in panels a-f to a STDDEV map for a time frame of $\pm$10min around the presented jet time. In panel h, the colour codding is the same as in panels a-f.}
    \label{AR_March_jet_length}
\end{figure*}

Step 4: The resulting STDDEV map was used to define the shape of the jet as follows. We determined the positions of pixels with the maximal value of STDDEV along the horizontal (X axis), vertical (Y axis), and both diagonal axes (from the top left to the bottom right and from the top right to the bottom left)  independently. We flagged these maximal values along each axis in a different colour. Figure \ref{AR_March_jet_length} (panels a-e) schematically presents this procedure in detail. 

Step 5: If at least two pixels flagged with the same colour are adjacent to each other, we call them a group. Our further analysis focused only on groups consisting of at least four pixels.  To increase the precision of the jet shape measurement, we gathered all groups ($\geq$  4 pixels) as a single map. The group concentrations indicated bright features in the solar atmosphere (e.g. a coronal loop or jet). Based on the visual inspection of AIA 193\AA~images, we decided which concentration of the flagged pixel groups indicated a jet. This group concentration corresponds to the main body of the jet.

Step 6: We fitted a quadratic polynomial along either the horizontal or vertical direction depending on the predominant orientation of the jet. This fitted curve does not contain any information on the beginning or end point of the jet but contains information on the shape. In a few complex and wide examples, it was impossible to obtain a good fit along the jet, and they were not included in the statistics.

Step 7: We analysed the intensity images of AIA 193\AA~for 20-minute time series. First, we calculated the average intensity for a single image, and we saved the result. A single image is represented by one value of the average intensity. We repeated the procedure for all images in the 20-minute time series. Next, we fitted a linear function to the average intensity versus time to find a trend of intensity changes.
Then, we subtracted a fitted trend of intensity from the 20-minute intensity time series. We obtained the 20-minute series corrected for slow ($>$20min) brightness changes unrelated to the jet. We call the obtained 20-minute time series `image trend corrected'.

Step 8: Then, we defined a background image as the trend-corrected image with the lowest average intensity obtained before the jet occurs. We smoothed a background image with a 2-pixel-wide Gaussian to reduce the noise. We subtracted this background from all images, which resulted in base difference images.

Step 9: For each base difference image in the range  $\pm$3min around the jet time, we computed the intensity profile along the curve obtained in step 5. For each intensity profile, we defined the longest continuous distance where the intensity is above 0 DN/s. At the ends of the continuous intensity, DN/s values equal to 0 DN/s define the jet beginning and end. The maximum continuous distance obtained from the whole time series is a jet length.

Step 10: Finally, we manually controlled the identification of the 0 DN/s points. Jets usually show a sharp increase in intensity at the footpoint and at the jet ends.

It should be noted that this length definition does not take projection effects into account. Length is only defined along orthogonal directions with respect to the viewing angle.

\subsection{Jet distribution velocity}\label{sec:ar_age} 
We only measured the jet propagation velocity of jets where the length measurement was possible. %
We used a polynomial curve obtained in step 6 of the jet length measurement procedure (Sect. \ref{section_DA_jet_length}) as the reference line to build a stack plot of intensity as a function of time and distance along the jet.
We determined the position of the pixel with the highest intensity at each time step in a stack plot.
The position of the pixels with the highest intensity usually creates a straight line pattern in the stack plot.
We fitted a straight line to the pixels with the highest intensity.
The slope of the fitted linear function defines the jet velocity propagation.

We successfully computed the velocity of 180 jets.
We neglected 59 jets because the influence of the surrounding plasma made a stack plot interpretation and line fitting ambiguous.

\subsection{The active region age}\label{sec:ar_age}

We determined the AR age to test whether this has any impact on the number of jets or their locations. We estimated the age using AIA 193\AA~observations and auxiliary HMI magnetograms.
We used JHelioviewer to track the AR evolution.
We identified visually, in the AIA 193\AA~channel, when the AR emerges.
We obtained the time of emergence with an accuracy of 3h.
We defined the AR age as the time between the AR emergence and the start of the observing time.
Moreover, all analysed ARs stay longer than 7 days after the end of observation; hence, we avoided the end phase of the AR life. 
The age of our five ARs is shown in Table~\ref{DA_table_AR_info}.

The AR age is difficult to define with high precision. For this work, only a qualitative comparison of the analysed ARs from three groups is needed: young (AR12737, AR12823), middle (AR12738, AR12822 ), and long-living (AR12803).
To properly determine the relation between the AR age and other physical parameters, a statistical analysis based on a larger sample of ARs is necessary.

\subsection{The magnetic field of the active regions}
The ARs were chosen to have different magnetic field morphologies and magnetic field strengths. We analysed the HMI LOS magnetogram at the beginning of each AR observation series (see Table~\ref{DA_table_AR_info}) to study the global magnetic field properties of the AR. 
We provided the analysis for the same field-of-view as presented in Figs.~\ref{AR_March_AIAJetflow} to~\ref{AR_May_JetflowS} for the middle observation time (see the lower-right corner of the HMI field-of-view in Figs.~\ref{AR_March_AIAJetflow} to~\ref{AR_May_JetflowS}). 
We calculated the magnetic flux ($\Phi$) based on the pixels above the noise level |B|>10G and the average strength of the unsigned magnetic field above |B|>100G to characterise the AR (see Table~\ref{DA_table_AR_info}). We arbitrarily chose a magnetic field limit of |B|>100G to avoid the influence of the weak magnetic field not related to ARs in the mean magnetic strength calculation.

Moreover, we defined the area of the AR where |B|>100G to compare the size of different ARs.
 Table~\ref{DA_table_AR_info} presents the area where |B|>100G for five ARs.

In the next sections we describe the results from the analysis of each AR separately. 

\section{Jets in AR12737}

\begin{figure}
    \centering
    \includegraphics[scale=0.67]{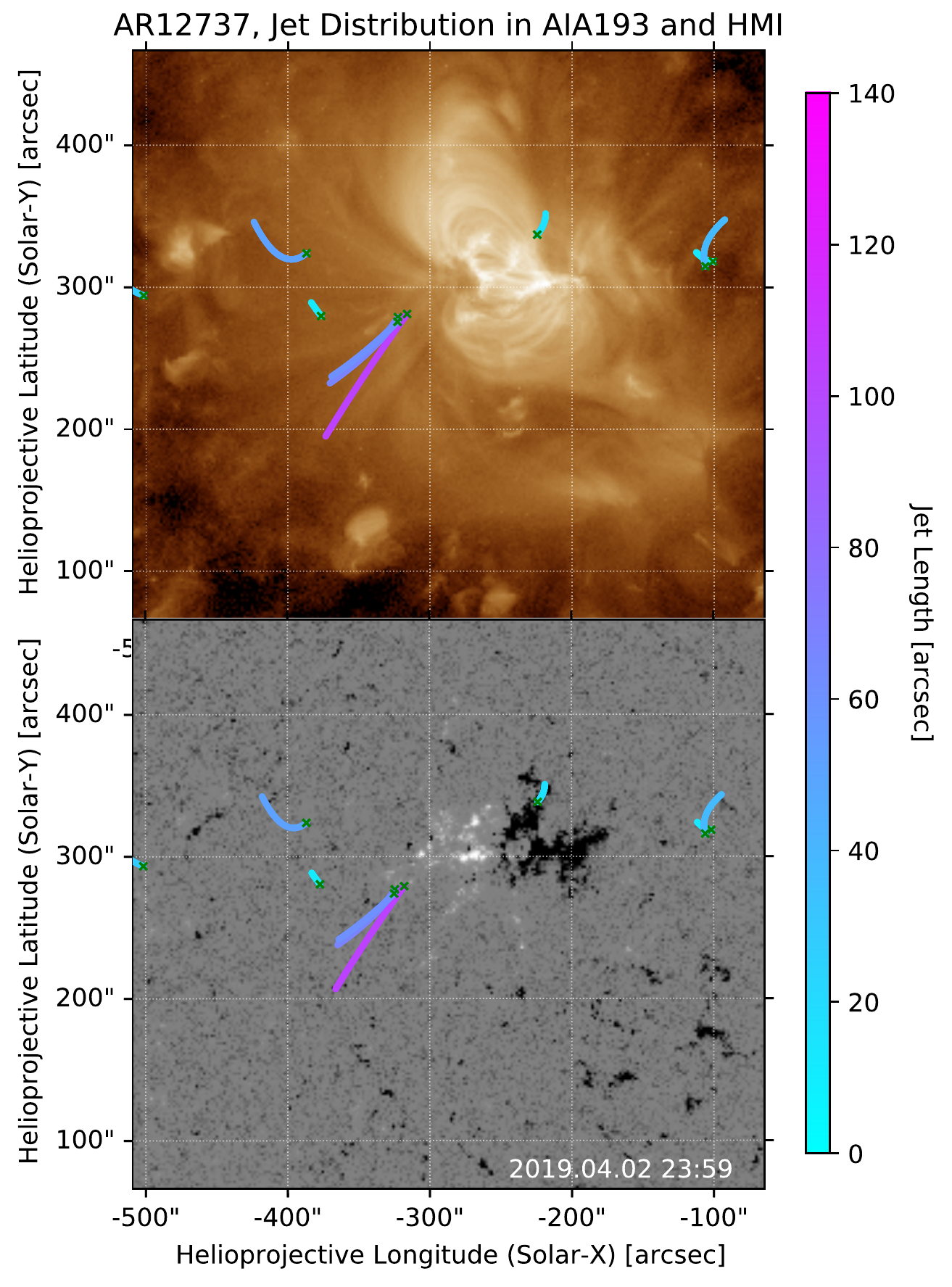}
    \caption{SDO observations of AR12737 with marked jet locations. The top figure shows the AIA 193 \AA\ image with the jet positions overlaid. The bottom figure shows the HMI magnetogram with the jet positions overlaid. The HMI magnetogram  range is [-300, 300] G. The length of the jet is indicated by the colour bar, and the footpoint is indicated by the green crosses. The figure shows the region of interest at the middle of the observation duration time.}
    \label{AR_March_AIAJetflow}
\end{figure}

We analysed the coronal images of AR12737 obtained between 00:00 UT 2~April~2019 and 
00:00 UT 4~April~2019. AR12737 was Hale class $\beta$ at the start of our observation. AR12737 had a bi-pole magnetic field, as seen in the HMI data in Fig. \ref{AR_March_AIAJetflow}. The leading negative magnetic field concentration was dominant, but dispersed.   Visual inspection of HMI data revealed  flux emergence and cancellation between the two dominant patches. In addition, small magnetic field patches cancelled each other out around the leading magnetic field. The observation time started three days after the first clear appearance of the AR in HMI data. AR12737 was a non-flaring AR, with no Geostationary Operational Environmental Satellite (GOES) classification flares noticed during our observation time.

The position and length of eight logged jets are shown in Fig. \ref{AR_March_AIAJetflow}. This AR produced the lowest number of jets compared with the rest of the investigated ARs.
One jet was rooted in the direct vicinity of the AR core, and the rest of the jets occurred at the AR border. All jets expanded in directions away from the AR core, and the footpoints were rooted closer to the centre of AR12737. The longest jets occurred in the region of expanded or `open' field lines at the eastern edge of the AR.

\section{Jets in AR12738}

We analysed images of the solar corona for AR12738 obtained between 00:00 UT 12 April~2019 and 12:00~UT 15~April~2019. AR12738 was Hale class $\alpha$ at the start of our observation. This region has the strongest magnetic field in our sample, with a clear leading negative sunspot.

\begin{figure}
    \centering
    \includegraphics[scale=0.48]{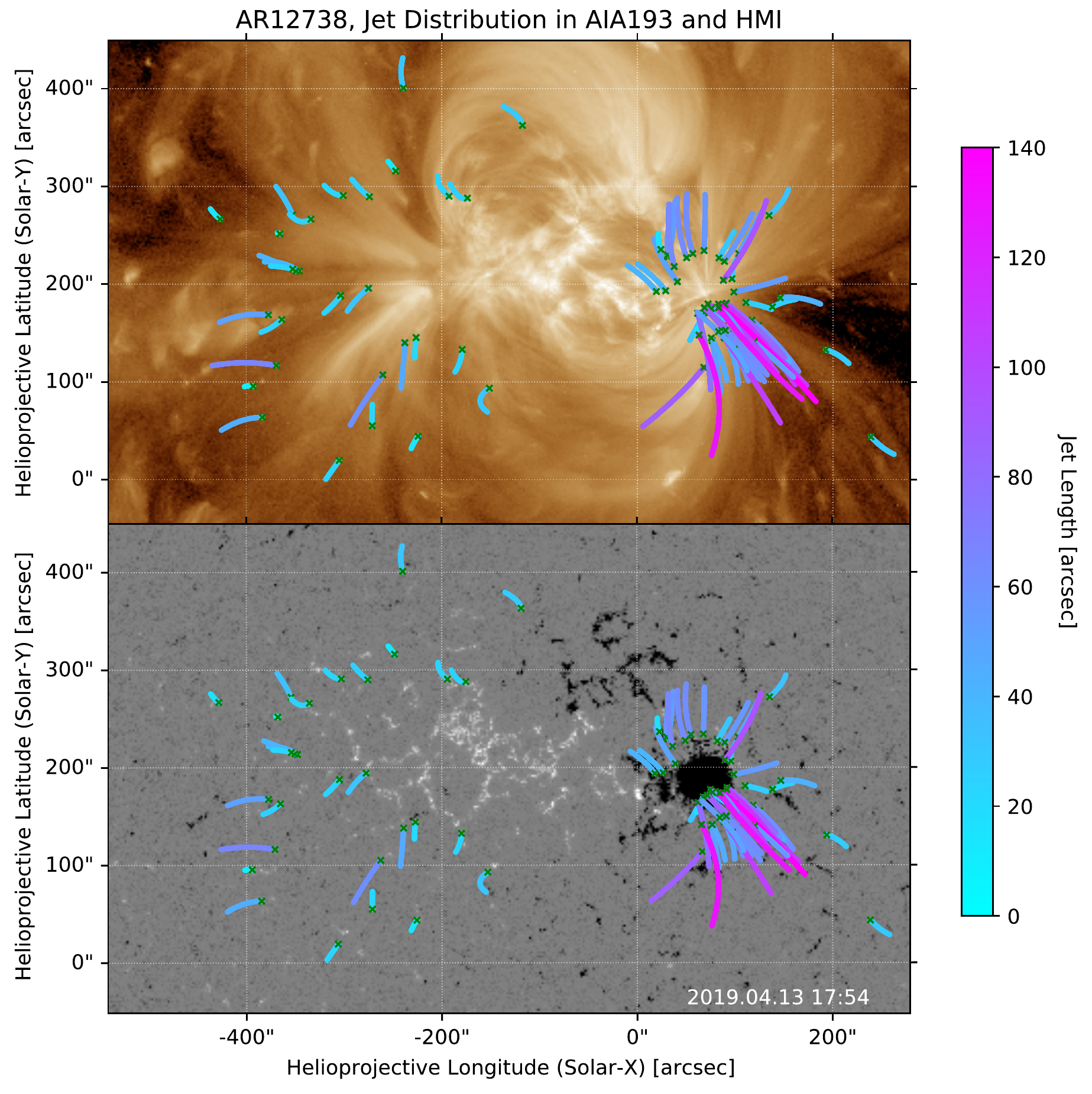}
    \caption{SDO observations of AR12738 with marked jet locations. The top figure shows the AIA 193 \AA\ image with the jet  positions overlaid. The bottom figure shows the HMI magnetogram with the jet positions overlaid. The HMI magnetogram  range is [-300, 300] G. The length of the jet is indicated by the colour bar, and the footpoint is indicated by the green crosses. The figure shows the region of interest at the middle of the observation duration time.}
    \label{AR_April_AIAJetflow}
\end{figure}
AR12738 had a bipolar magnetic field structure, as seen in Fig. \ref{AR_April_AIAJetflow}. The leading part of the AR has a negative magnetic field. The positive lagging magnetic field is significantly dispersed. To the north-east of the leading negative spot is a secondary dispersed negative field region. It lies close to the positive patch and shows evidence for strong flux cancellation at the closest position. The leading negative magnetic field is surrounded by small negative and positive magnetic field patches, which showed flux cancellation upon visual inspection. 

In AR12738, we found 92 jets over three and a half days. There are two dominant jet-producing regions, as shown in Fig. \ref{AR_April_AIAJetflow}.
The first jet-producing region (which we refer to as jet-producing Region I)  is around the leading negative magnetic field of AR12738 (80$^{\prime\prime}$,~200$^{\prime\prime}$). The mean magnetic field of the leading magnetic field patches is around -600 G.  Most of the jets rooted in Region I emanate from the north or south-west directions. There were several jets produced at the same location in the south-western part of Region I. They had a very long (up to 130$^{\prime\prime}$) and narrow shape. They had a footpoint rooted slightly inside the dominant magnetic field patch. However, the footpoint location inside the dominant magnetic field patches can be related to projection effects. While magnetograms show photospheric magnetic fields, jet footprints with a coronal temperature are located higher up. The difference in height could cause a small apparent offset in the location, making the jet footpoints appear to be located inside the one-polarity region.
The northern part of Region I produced fewer jets than the southern. However, the jets produced in the northern part of Region I had a more complex shape. All jets in Region I expanded outwards from the centre
of the circular negative magnetic field patch. The footpoints of the jets determined from AIA~193\AA~images encircled the leading magnetic field patch. 
The second jet-producing region (Region II) is around the lagging positive magnetic field region. The average magnetic field of Region II is 45 G, but the individual patches have a mean magnetic field of around 500G. The jets rooted in Region II  are in dispersed locations, rarely in the vicinity of one another. They were oriented away from the centre of the positive magnetic field patch.

All large jets rooted in Region I expanded in northern or southern directions. The seven largest jets (above 100$^{\prime\prime}$) in AR12738 are oriented towards the south. The jet size was smaller in Region II, where the longest  jet had a length of less than 70$^{\prime\prime}$. The leading magnetic polarity (Region I) had the longest jets, and it had more frequent jets and a stronger magnetic field (-600 G) compared with Region II (45 G).

\section{Jets in AR12803}
We analysed coronal images of AR12803 obtained between 12:00 UT 23 February 2021 and 12:00 UT 25 February 2021. AR12803 was not classified with a Hale classification at the start of our observation. More information on the AR and its surroundings is given in Table \ref{DA_table_AR_info}.

\begin{figure}
   \includegraphics[scale=0.58]{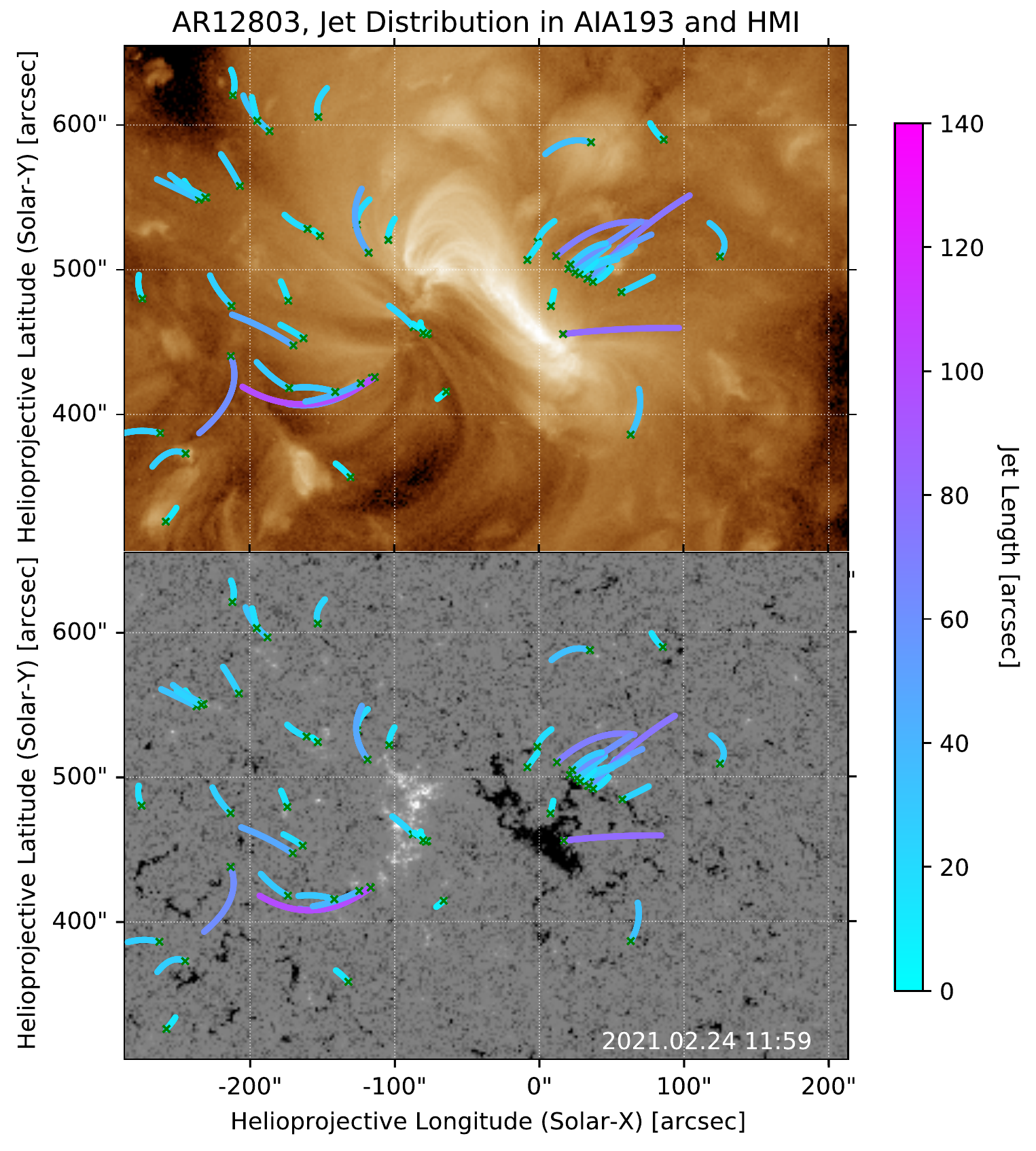}
    \caption{SDO observations of AR12803 with marked jet locations. The top figure shows the AIA 193 \AA\ image with the jet  positions overlaid. The bottom figure shows the HMI magnetogram with the jet positions overlaid. The HMI magnetogram  range is [-80, 80] G. The length of the jet is indicated by the colour bar, and the footpoint is indicated by the green crosses. The figure shows the region of interest at the middle of the observation duration time.}
              \label{AR_jetlength_Feb}%
    \end{figure}

AR12803 is shown in Fig.~\ref{AR_jetlength_Feb}. No significant changes in the coronal loops were observed during the observation period. 
AR12803 had a bipolar magnetic structure with both the leading and lagging magnetic patches being dispersed. The leading negative patch shows small and highly dynamic positive parasitic patches. 

There were 70 jets distributed in two dominant jet-producing regions. One was a compact region on the north-western edge of the leading magnetic patch. This region produced several jets at the same location while also being the major producer of the longest jets in AR12803. The jets were related to a magnetic flux cancellation region. This region produced more jets than the rest of the AR for approximately ten hours.

Several small flux cancellation areas throughout the positive magnetic patch were related to jets.  One significantly larger jet ($\approx 100''$) appeared on the southern edge, where stronger magnetic flux emergence and cancellation occurred.

\section{Jets in AR12822}
        
We analysed images of the solar corona for AR12822 obtained between 12:00 UT 12 May 2021 and 12:00 UT 14 May 2021. AR12823 was Hale class $\beta$-$\delta$ at the start of our observation. More information on the AR and its surroundings is given in Table \ref{DA_table_AR_info}.

\begin{figure}
   \includegraphics[scale=0.62]{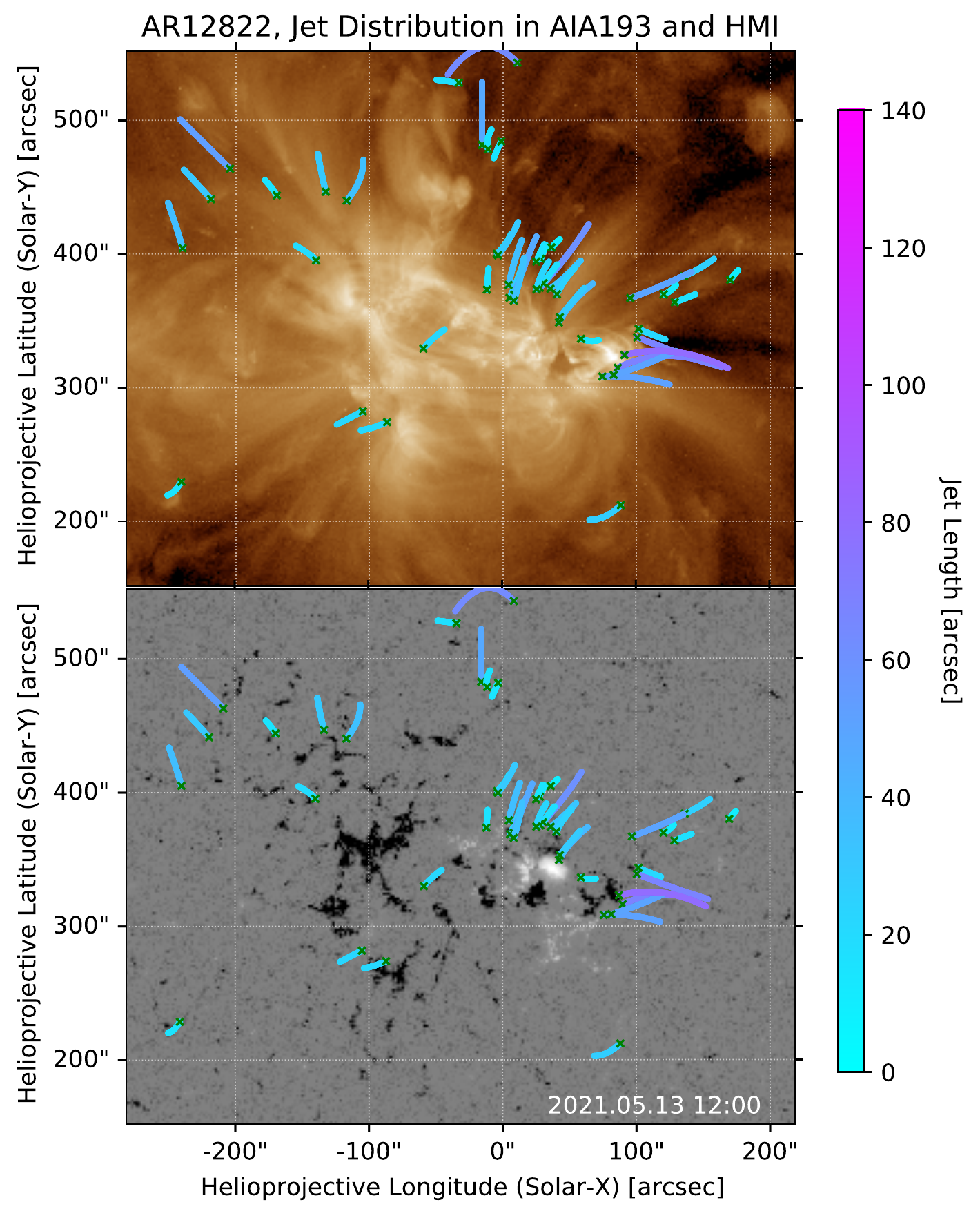}
    \caption{SDO observations of AR12822 with marked jet locations. The top figure shows the AIA 193 \AA\ image with the jet  positions overlaid. The bottom figure shows the HMI magnetogram with the jet positions overlaid. The HMI magnetogram  range is [-80, 80] G. The length of the jet is indicated by the colour bar, and the footpoint is indicated by the green crosses. The figure shows the region of interest at the middle of the observation duration time.}
    \label{Ar_May_Jetflow}%
    \end{figure}

We investigate AR12822 as seen in AIA data in Fig.~\ref{Ar_May_Jetflow}. The northern coronal loops strongly evolved, starting at approximately 16:00~UT 13~May~2021. At this point, new coronal loops emerged, rooted at the top end of the western fan loop region. Small loop brightenings preceded this change. Several smaller loop brightenings appeared on the leading edge of AR12822, oriented towards the south-east.

Figure~\ref{Ar_May_Jetflow} shows the SDO/AIA magnetogram of AR12822. Both magnetic field polarities are strongly dispersed. However, the negative magnetic field is stronger and more dispersed than the positive one.

There were 50 jets distributed mainly in two dominant regions. One was located at a short distance from the leading positive magnetic field concentration and the second on the north-eastern edge of the lagging negative field patch. Most jets  corresponded to small-scale magnetic field concentrations. Moreover, the jets avoided the major magnetic field patches. The jets were oriented away from the AR, with the footpoints (viewed in AIA 193~\AA) being closer to the nearest dominant magnetic field patch.

\section{Jets in AR12823}

We analysed images of the solar corona for AR12823 obtained between 12:00 UT 12 May 2021 and 12:00 UT 14 May 2021. AR12823 was Hale class $\beta$ at the start of our observation. More information on the AR and its surroundings is given in Table \ref{DA_table_AR_info}.

\begin{figure}
   \includegraphics[scale=0.68]{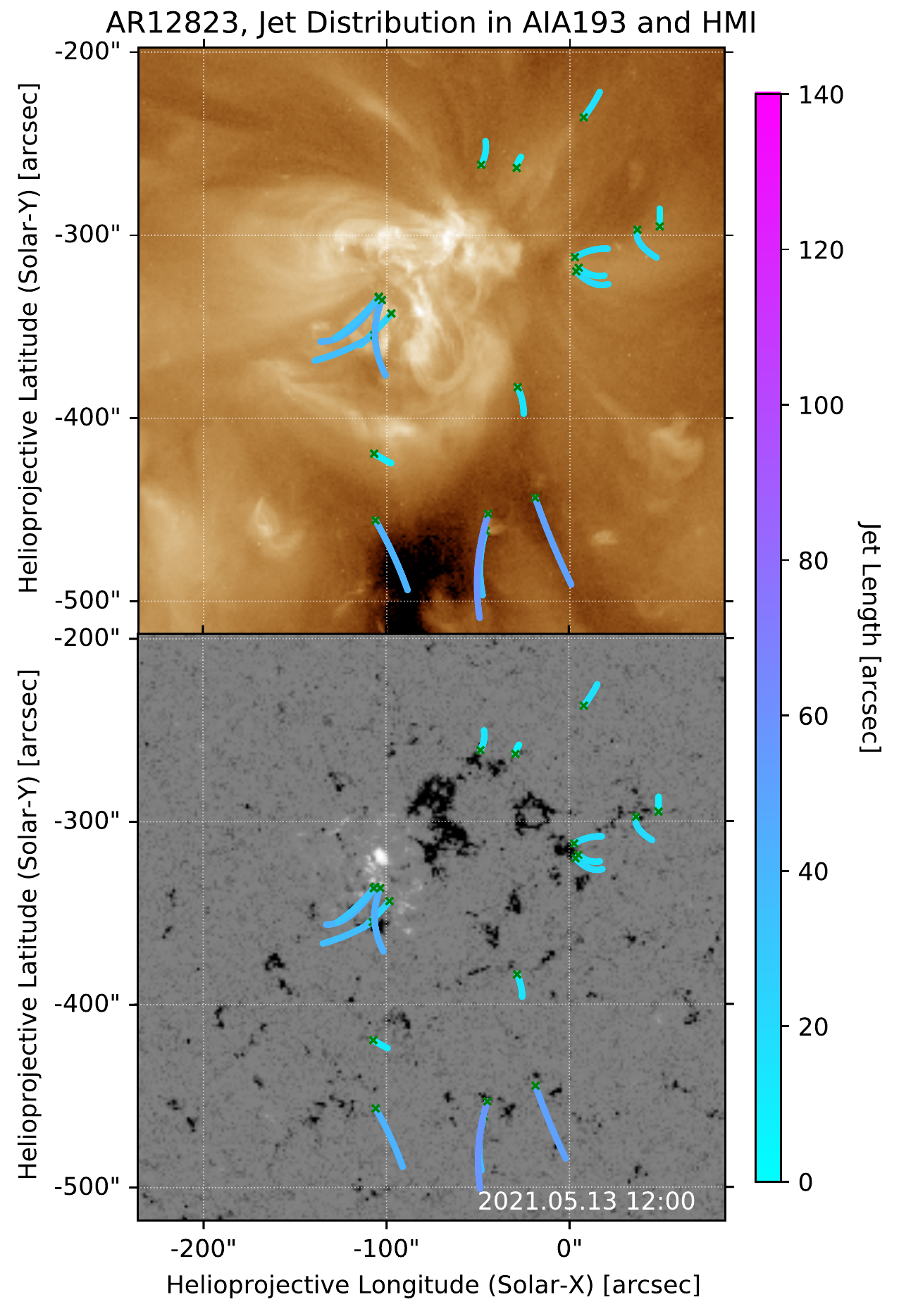}
   \caption{SDO observations of AR12823 with marked jet locations. The top figure shows the AIA 193 \AA\ image with the jet  positions overlaid. The bottom figure shows the HMI magnetogram with the jet positions overlaid. The HMI magnetogram  range is [-80, 80] G. The length of the jet is indicated by the colour bar, and the footpoint is indicated by the green crosses. The figure shows the region of interest at the middle of the observation duration time.}
              \label{AR_May_JetflowS}%
    \end{figure}

We investigated AR12823 as presented in Fig.~\ref{AR_May_JetflowS}.  AR12823 had a less defined coronal loop structure than the other  ARs (e.g. AR12738) and showed coronal loops opening up and re-emerging. At the beginning of the observation, it had coronal loops from the centre towards the north-east. There was an increasing number of visible coronal loops at the northern end of the AR, originating at the leading magnetic field patch. 

Figure~\ref{AR_May_JetflowS} presents the SDO/HMI magnetogram of AR12823. It had a dispersed negative magnetic field patch to the west and a smaller, more compact and positive magnetic field patch to the east.   During the time of intense loop brightening, no significant structural changes appeared in the magnetic field. 

 There were 19 jets distributed in three locations. The first is located next to a fan loop region north-east of the centre. The second jet production region is located very slightly south of the centre and contains five jets. The last location is south of the AR, near dominantly negative magnetic field patches. All jets are oriented away from AR12823.

    \begin{figure}
    \includegraphics[scale=0.71]{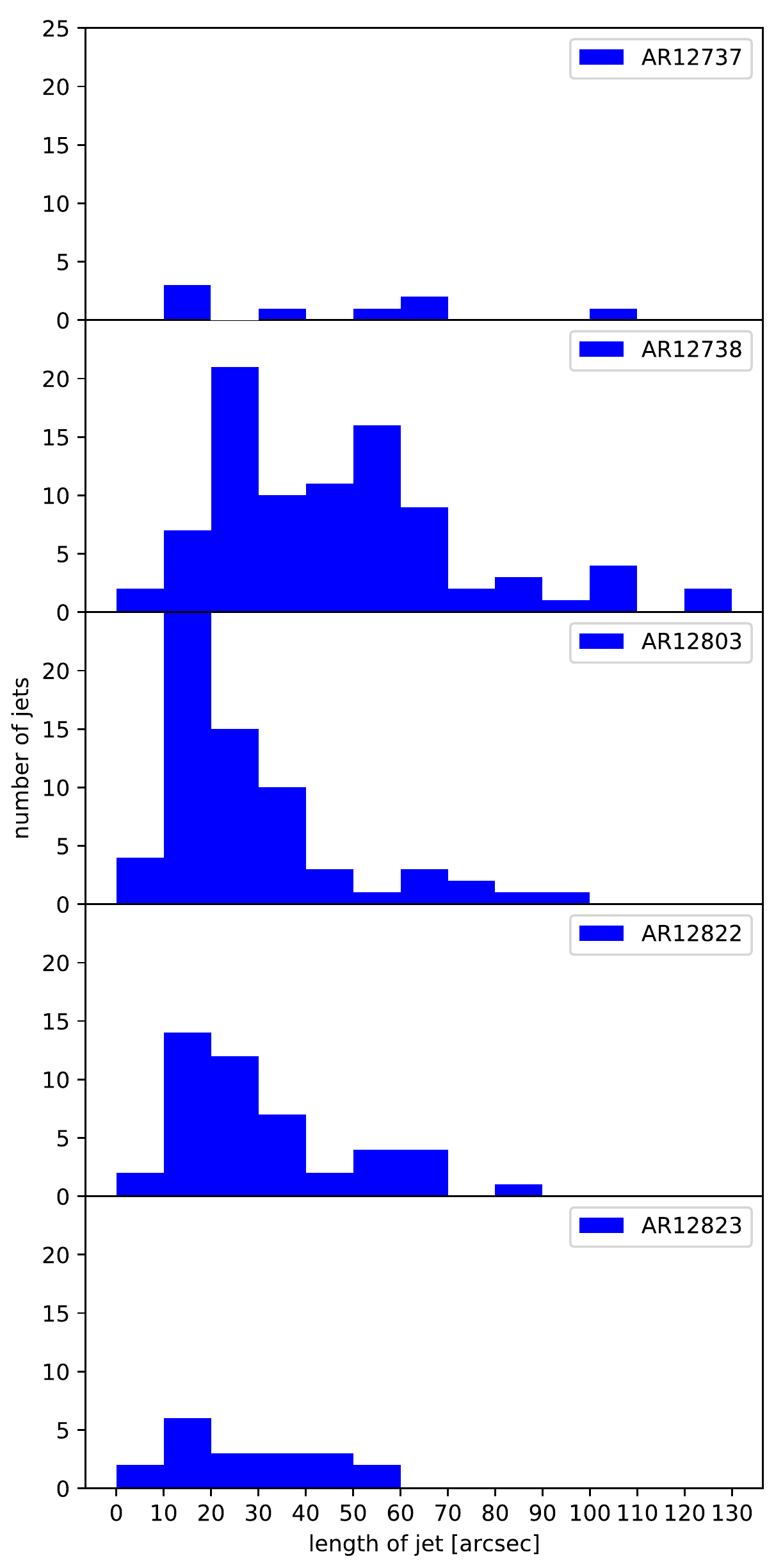}
    \caption{Histogram of the measured jet lengths for five ARs.}
    \label{jet_length_summary}%
\end{figure}



  
\section{Conclusions}

  We have analysed five ARs to explore the occurrence of jets as well as their location within the AR. In all cases, the ARs showed jets and there were few jets in the central core of the ARs. This could be because of the high intensity in this region, or it could be physically due to the fact that jets have a preference for regions that can have an interaction between closed and open or extended magnetic fields. This study exclusively focused on jets that have a clear route of escape (i.e. not within closed loop structures).  Most jets were located at the edges of the magnetic field polarities regardless of whether they were strong compact regions or weak dispersed regions. The strongest and longest jets were seen at the strongest leading polarities.
  
 We find that the number of the jets per time unit in an AR depends on the age of the AR.
 The young ARs (e.g. AR 12737 and AR12823) showed significantly fewer jets than the old ARs (e.g. AR12803). We only looked at five ARs in this work, and the relationship between the number of jets and the AR age needs further investigation via a statistical study.
 
 Our analysis demonstrates that the number of jets per time  in an AR is independent of the global magnetic field properties of the AR. 
 The analysis, based on five ARs, suggests that the number of jets per time unit is independent of
 (i) the average unsigned magnetic field of ARs, (ii) the total magnetic field flux, and (iii) the AR area.

Our study suggests that
longer jets occur in older ARs (Fig.\ref{jet_length_summary}). The jet length is independent of the average magnetic field of the AR and the area of the AR.

\begin{figure}
    \includegraphics[scale=0.71]{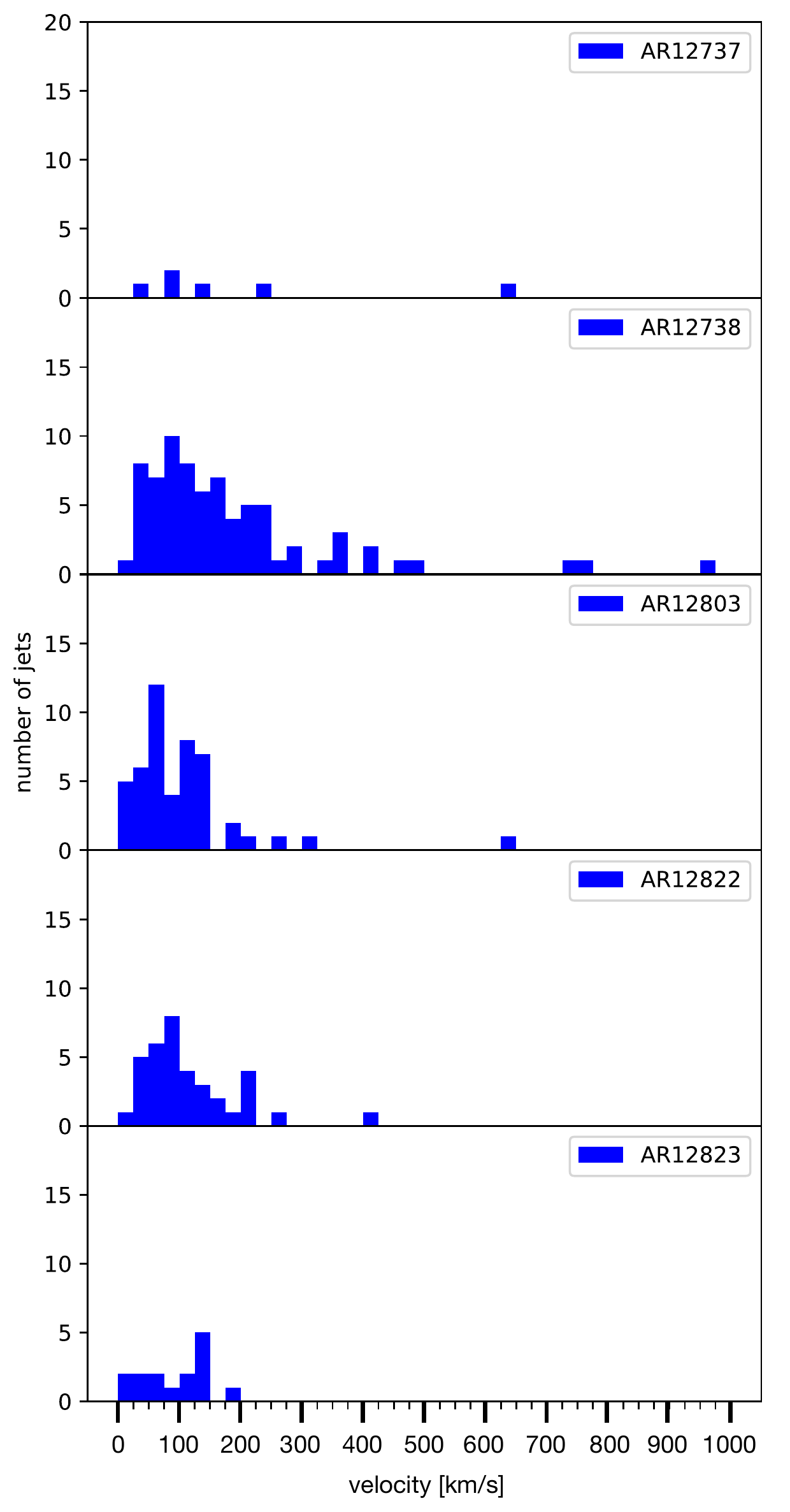}
    \caption{Histogram of the measured jet velocity propagation for five ARs.}
    \label{histogram_velocity}%
\end{figure}

Our analysis demonstrates that most of the jets have a velocity in the range 0 to 200 km s$^{-1}$ (Fig.~\ref{histogram_velocity}).
Jets with velocities higher than 400 km s$^{-1}$ are rare.
The median velocity is 114 km s$^{-1}$.
Our velocity measurements are consistent with previous studies.
 In conclusion, we find that all our ARs, regardless of how weak the magnetic field was, produced jets. The jets were found to be at the edges of the magnetic flux, again regardless of how strong or how dispersed the magnetic field was. The longest jets were found in the region with the strongest magnetic field -- in particular, in the leading magnetic polarity. There are indications that the age of the AR plays a role in jet productivity. 
  
  Our study is the first step towards enhancing our understanding of the distribution of the coronal jets in ARs. A future statistical study is required to address questions that emerged from this study, in particular the questions of: how the jet distribution changes during the AR evolution; whether the average size of the jets depends on  the jet location in an AR; whether the number of jets per day in ARs changes with the solar activity cycle; whether the vicinity of the AR (e.g. other ARs, quiet Sun regions, or coronal hole regions) influences the number of observed jets; whether the jet duration and velocity propagation depend on the jet location in ARs; how the distribution in the AR of the jet's length and velocity is measured along the jet propagation direction in 3D (non-projected); and, finally, what the spatial jet distribution is in multi-wavelength observations (e.g. in different AIA channels). Further study and data obtained with simultaneous high-resolution observations (e.g. from the SDO and Solar Orbiter) are highly desirable to answer these questions.
 
\begin{acknowledgements}
This work is supported by Swiss National Science Foundation - SNF. SDO data  are  courtesy of NASA/SDO and the AIA, EVE, and HMI science teams. 
\end{acknowledgements}

%
%
\bibliographystyle{aa} 
   \bibliography{references.bib} 
%



\end{document}